\documentclass[aps,prl,twocolumn,showpacs,showkeys,groupedaddress]{revtex4-2}

\usepackage{graphicx}
\usepackage{rotating}
\usepackage{amssymb}
\usepackage{amsmath}
\usepackage{epsfig}
\usepackage[normalem]{ulem}
\usepackage{bm}
\usepackage{color}
\usepackage{subfigure}
\usepackage[hidelinks]{hyperref}

\date{\today}

\begin{document}

\author{Carlos E. P. Abreu}
\affiliation{Departamento de Física, Instituto de Geociências e Ciências Exatas,
Universidade Estadual Paulista, UNESP, 13506-900, Campus Rio Claro, São Paulo, Brasil}
\affiliation {Instituto Federal de Educação, Ciência e Tecnologia do Sul de Minas Gerais - IFSULDEMINAS, Três Corações, 37417-158, Minas Gerais, Brasil} 

\author{Joelson D. V. Hermes}

\affiliation {Instituto Federal de Educação, Ciência e Tecnologia do Sul de Minas Gerais - IFSULDEMINAS, Inconfidentes, 37576-000, Minas Gerais, Brasil}

\author{Diogo Ricardo da Costa}
\affiliation{Departamento de Física, Instituto de Geociências e Ciências Exatas,
Universidade Estadual Paulista, UNESP, 13506-900, Campus Rio Claro, São Paulo, Brasil}

\author{Everton S. Medeiros}
\affiliation{Institute for Chemistry and Biology of the Marine Environment, Carl von Ossietzky University Oldenburg, 26111 Oldenburg, Germany}

\author{Rene O. Medrano-T}
 \email{rene.medrano@unifesp.br}
 \affiliation{Departamento de Física, Universidade Federal de São Paulo,UNIFESP, 09913-030, Campus Diadema, São Paulo, Brasil}
\affiliation{Departamento de Física, Instituto de Geociências e Ciências Exatas,
Universidade Estadual Paulista, UNESP, 13506-900, Campus Rio Claro, São Paulo, Brasil}

\title{Singular fractal dimension at periodicity cascades in parameters spaces }

\begin{abstract}
In the parameter spaces of nonlinear dynamical systems, we investigate the boundaries between periodicity and chaos and unveil the existence of fractal sets characterized by a singular fractal dimension. This dimension stands out from the typical fractal dimensions previously considered universal for these parameter boundaries. We show that the singular fractal sets dwell along parameter curves, called extreme curves, that intersect periodicity cascades at their center of stability in all scales of parameters spaces. 
The results reported here are generally demonstrated for the class of one-dimensional maps with at least two control parameters, generalizations to other classes of systems are possible.
\end{abstract}

\maketitle

In nonlinear systems, order and chaos are two profoundly contrasting dynamics, yet they often intricately intertwine within the system's parameter spaces. 
Although, the domains of parameter sets corresponding to chaotic attractors in general are not continuous, they are dense enough to present positive Lebesgue measure \cite{Jakobson1981}. Arbitrarily close to these sets there are continuous periodic windows with stable periodic behavior \cite{Graczyk1997}. Consequently, the parameters that lead to chaotic attractors form {\it fat fractal} sets, and chaotic dynamics can be replaced by stable periodic behavior through an arbitrarily small variation of the system parameters \cite{hunt1997structure}. Moreover, since the periodic windows occur across various scales of the parameter space, they give rise to periodicity {\it cascades}, an infinite set of self-similar periodic windows, densely distributed in the parameter space. Along cascades, the distribution of periodic windows can be governed by scaling rules related to their size \cite{Yorke1985}, period \cite{Kaneko1982}, and other topological measures associated with their periodic orbits \cite{Englisch1991,Medeiros2013}.

In the literature, considerable efforts have been dedicated to characterizing the complexity arising from the self-similarity of periodic windows within the context of one-dimensional maps, where only a single bifurcation parameter is available. Grebogi et al. developed an approach to determine the ``exterior dimension'' \cite{grebogi1985exterior} of these sets through an estimation of the so-called ``uncertainty exponent'' \cite{grebogi1983final}. They obtained an estimated value of this exponent for the periodic windows of the quadratic map, which was found to be $\alpha=0.413(5)$. Additionally, Farmer proposed an alternative approach to also characterize the intertwined structure of periodicity and chaos, finding scaling exponents to be $\beta=0.45(4)$ for the quadratic and sine maps \cite{farmer1985sensitive}. He conjectured that this exponent could be universal among one-dimensional maps up to a certain order of their maxima. Subsequently, Hunt et al. \cite{hunt1997structure} theoretically and numerically estimated the uncertainty exponent solely for ``large'' chaotic attractors of the quadratic map, i.e., excluding the ``small'' chaotic attractor appearing via the Feigenbaum scenario. 
This approach yields a different value for the uncertainty exponent, $\gamma=0.51(3)$. Such discrepancy between $\gamma$ and $\alpha$ has been addressed by Joglekar et al., who demonstrated a relationship between these exponents and conjectured that both values are universal for one-dimensional maps with a quadratic maximum \cite{joglekar2014scaling}.

In planar parameters spaces, where two bifurcation parameters are available, cascades of self-similar periodic windows can manifest themselves in two main ways: 
i) Aligned towards specific directions of two-parameter spaces, giving rise to periodicity cascades that accumulate into a parameter region corresponding to periodic behavior \cite{Bonatto2007,Bonatto2008,Medeiros2013}. 
ii) Periodicity hubs, characterized by infinitely many spiral-shaped sequences emanating from a single point in the parameter space corresponding to a homoclinic bifurcation \cite{Vitolo2011,Barrio2011,Barrio2012}. Such nontrivial organization of periodic windows has been found in two-parameter spaces of several classes of dynamical systems in either computational \cite{gallas1993structure, Bonnato2005, Stoop2012, Medrano2014, Rocha2015, Maranhao2016, HegedHus2018, Raphaldini2021, Hallier2022} and laboratory experiments \cite{maranhao2008experimental, Stoop2010, Viana2010}. In this context, Medeiros et al. estimated the uncertainty exponent $\alpha$ for three different continuous-time systems containing periodicity cascades in their planar parameter spaces  \cite{medeiros2017sensitive}. They found the values of $\alpha$ for all systems to be in the interval $\alpha=0.40(4)$, corroborating the conjectured universality of this exponent even in another class of dynamical systems \cite{medeiros2017sensitive}.

Here, we present evidence challenging the longstanding belief in this universality. For that, we consider the class of one-dimensional maps generically governed by the equation:
\begin{equation}
\label{Eq:one_dim_map}
 x_{n+1}=f(x_n,{\bm a}),
\end{equation}
where $f$ is sufficiently smooth and at least bimodal, the variable $x \in \mathbb{R}$ represent the states of the system, and the vector ${\bm a} \in \mathbb{R}^N$ accounts for the $N \geqslant 2$ control parameters. In these high-dimensional parameter spaces, we first specify parameter curves, called extreme curves, intersecting periodicity cascades at their center of stability. 
Subsequently, we estimate the uncertainty exponents along the extreme curves and obtain $\alpha \sim 0.23$.
Interestingly, we find that the transition from these values to the ones previously found in the literature occurs abruptly in the vicinity of the extreme curves. Moreover, by considering the uncertainty exponent as an approximation of the exterior codimension, we obtain the singular fractal dimension of parameter sets dwelling at periodicity cascades and chaos boundaries.

{\bf Extreme curves:} Since our proposed singular fractal sets dwell in the multidimensional parameter space of mappings in Eq.~(\ref{Eq:one_dim_map}), we now specify the location of such fractal objects. In general, starting at an initial condition $x_0$, the successive iterates of Eq.~(\ref{Eq:one_dim_map}) yields trajectories that, depending on the parameters ${\bm a}$, eventually approach an asymptotic solution such as a fixed point, a periodic orbit, or a chaotic attractor. 
However, even before converging, such trajectories may contain sequences of critical points $\{x_{1}^{\ast}, x_{2}^{\ast}, \dots, x_{i}^{\ast}, \dots \}$, each one satisfying $f'(x_{i}^{\ast})=0$, associated to extreme points (local maxima or minima) of the mapping in Eq.~(\ref{Eq:one_dim_map}). Naturally, any pair of critical points, $x_{i}^{\ast}$ and $x_{j}^{\ast}$, may be connected by $k$ successive iterates of the mapping, i.e., $x_j^{\ast}=f^k(x_i^{\ast})$, with $k \in \mathbb{N}$, $f^0 = $ identity and $f\equiv f^1$. Such trajectories of length $k$ connecting critical points in the state space of the mapping are referred to as {\it $k$-extreme orbits} and, their corresponding parameters $e^{ij}_k=\{ {\bm a} \in \mathbb{R}^N | x_j^{\ast} = f^{k}(x_i^{\ast},{\bm a}) \}$, constitute codimensional one sets known as {\it extreme curves} in planar parameter spaces \cite{da2016role}. The extreme curves host the singular fractal sets of interest in this letter.

Now, we consider an explicit one-dimensional mapping with two or more control parameters to demonstrate the existence of extreme curves in their parameter space. For that, we first employ the so-called Logistic-Gauss map, as defined by the following function \cite{da2021logistic}:
\begin{equation} \label{Eq:log_gauss}
     f(x,a_{1},a_{2},a_{3})=\exp{\{{-a_1 [x(1-x^{a_{2}})]^{2}}\}}+a_{3},
\end{equation}
with $a_2 \in \mathbb{R_+^*}$. 
To investigate the fractal dimension of the extreme curves $e^{ij}_{k}$, we define a two-dimensional cross-section $\Gamma=\mathbb{R}\times \mathbb{R_+^*}$, by keeping $a_3=0.1$ constant. 
Hence, from Eq.~(\ref{Eq:log_gauss}), we first obtain the sequence of critical points:
\begin{equation} \label{Eq:critical}
     x_{1}^{\ast}=0, \quad
     x_{2}^{\ast}= (1+a_2)^{-\frac{1}{a_2}}, \quad \mathrm{and} \quad 
     x_{3}^{\ast}=1.
\end{equation}
If $a_2$ is a rational number with even numerator, two more critial points come out, $x_4 = -x_2$ and $x_5 = -x_3$.
Next, by solving the equation 
$x_{3}^{\ast}=f(x_{2}^{\ast},{\bm a})$, it is obtained the extreme curve $e^{23}_{1}$:

\begin{equation}  \label{Eq:extreme1}
    a_1 = -\frac{\ln(0.9)(a_2+1)^{2+2/a_2}}{a_2^2}.
\end{equation}
Similarly, we obtain the function of $e^{32}_{2}$:
\begin{equation} \label{Eq:extreme2}
    a_1=-\frac{\ln[(1+a_2)^{-\frac{1}{a_2}}-0.1]}{1.21\,(1-1.1^{a_2})^{2}}.
\end{equation}

\begin{figure}[h]
\centering
\includegraphics[width=.48\textwidth]{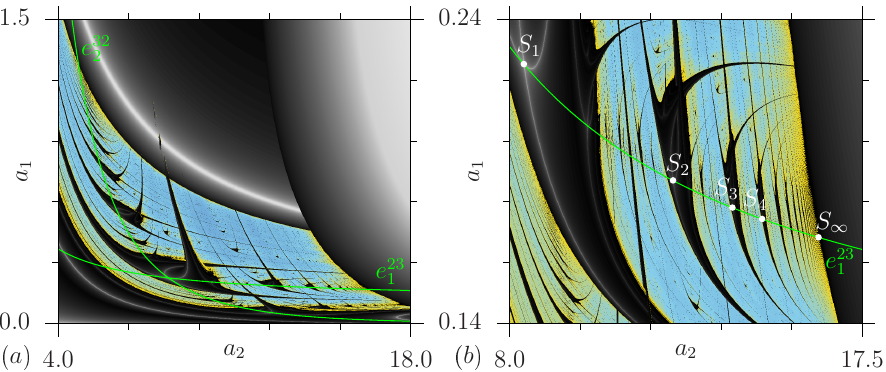}
\caption{(a) Periodicity cascade in the parameter plane $\Gamma$ along the green extreme curves $e_1^{23}$ and $e_2^{32}$. Dark shading represents periodic dynamics ($\lambda<0$), while yellow-blue shadings indicate chaotic behavior ($\lambda>0$). (b) Emphasis on $e_1^{23}$ crossing the cascade through all superstability locus, sequence $(S_i)_{i=1}^\infty$.}
\label{figure_1}
\end{figure}

To associate the extreme curves with the periodicity cascades occurring in $\Gamma$, for each parameter pair ($a_1,a_2$), we estimate the Lyapunov exponent of the mapping prescribed in Eq.~(\ref{Eq:log_gauss}) as $\lambda=\lim_{n \rightarrow \infty}\frac{1}{n}\sum_{i=1}^{n}\ln|f'(x_i)|$. Hence, in Fig.~\ref{figure_1}, we show $\Gamma$ with the color code standing for the amplitude of $\lambda$. 
Darker shadings indicate periodic ($\lambda<0$), and yellow-blue shadings indicate chaotic dynamics ($\lambda> 0$). 
The extreme curves $e^{23}_{1}$ and $e^{32}_{2}$ are traced in green in panel (a). 
In (b) we highlight a cascade of periodic windows relative to $e^{23}_{1}$. 
Note that this curve crosses over all periodic windows intersecting points corresponding to {\it superstability locus}, local where two white superstable curves are crossing \cite{barreto1997high,Faccanha2013}.
Notably, if a stable periodic orbit contains extreme orbits in its extension, it automatically fits the criteria for superstability locus.
These extreme and superstable orbits \footnote{For superstability we mean the parameter set in which stable periodic orbits exert faster attractiveness over nearby trajectories, i.e., when $x_i^*$ is part of a periodic orbit.} populate the superstability locus $S_1, S_2,\dots,S_{\infty}$ along the extreme curve $e^{23}_{1}$ throughout this sequence of periodic windows. 
Moreover, the curve $e^{23}_{1}$ also intersects the superstability locus of periodic windows composing smaller subsets accumulating at each periodic window.
The infinitely many   {subsets} 
of periodic orbits, occurring in all scales of cross-section $\Gamma$, traversed by the curve $e^{23}_{1}$, forms the periodicity cascade.
The same is for $e_2^{32}$.

{\bf Exterior dimension:} The self-similar organization of the periodicity cascades visualized in Fig.~\ref{figure_1} points to the existence of fractal sets embedded in the parameter cross-section $\Gamma$. However, the parameter sets leading to either chaotic or periodic behavior possess nonzero volume. Therefore, their dimension is the same as the Euclidean dimension of the embedding space, i.e., an integer. 
To characterize the scaling of the cascades, different approaches have been proposed in the literature \cite{grebogi1985exterior, farmer1985sensitive,hunt1997structure}.
We focus on the framework of {\it exterior dimension} presented in \cite{grebogi1985exterior} in which the scaling observed in Fig.~\ref{figure_1} is attributed to a fractal geometry of the boundaries between the parameters sets leading to periodic or chaotic behavior. 
In our discussion, this fractality reflects the exterior dimension of the chaotic fat fractal set along specific curves.
To better introduce this concept, consider a region $\mathcal{B}$ of the chaotic set and extend this region by an arbitrary small amount $\varepsilon$. Call this extended region as $\mathcal{B}(\varepsilon)$.
The exterior dimension $d_x$ is given by \cite{grebogi1985exterior}:
\begin{equation}
d_x=D-\lim_{\varepsilon \rightarrow 0}\frac{\ln V[\mathcal{B}(\varepsilon)-\mathcal{B}]}{\ln \varepsilon},
\label{Eq:codimension}
\end{equation}
where $D$ is the unitary dimension of the curves and $V[\mathcal{B}(\varepsilon)-\mathcal{B}]$ is the remaining ({\it exterior}) volume excluding the volume of the original set $\mathcal{B}$.
In accordance with \cite{grebogi1985exterior},
$V[\mathcal{B}(\varepsilon)-\mathcal{B}]$ 
can be estimated by computing the fraction of parameters $f(\varepsilon)$ that are uncertain over $\varepsilon$-size perturbations along specific curves of the cross-section $\Gamma$. 
More specifically, we first choose a point $\varphi = (a_1, a_2)$ at random on the extreme curve of interest [Eqs. (\ref{Eq:extreme1}) or (\ref{Eq:extreme2})] and perturb $\varphi$ by considering an distance $\varepsilon$ in both direction of the curve to obtain the points $\varphi_-$ and $\varphi_+$.
Subsequently, we evaluate the Lyapunov exponent $\lambda$ for each parameter value $\varphi_-$, $\varphi$, and $\varphi_+$. If in these three parameter we identify chaotic $(\lambda>0)$ and periodic $(\lambda<0)$ attractors we record the central parameter $\varphi$ as “uncertain” for the particular value of the perturbation $\varepsilon$. Otherwise, $\varphi$ is said to be ``certain" and is disregarded.
We repeat this procedure for a large number of $\varphi$ values and calculate the fraction of uncertain parameters $f(\varepsilon)$ for each perturbation $\varepsilon$ in the interval $10^{-12}\le \varepsilon \le 10^{-3}$. The fraction $f(\varepsilon)$ is known to depend on the perturbation $\varepsilon$ as a power-law, $f(\varepsilon)\sim \varepsilon^{\alpha}$, which the exponent $\alpha$ is simply $\alpha=\lim_{\varepsilon \rightarrow 0}\frac{\ln f(\varepsilon)}{\ln \varepsilon}$. 
Therefore, since $f(\varepsilon)$ approaches to $V[\mathcal{B}(\varepsilon)-\mathcal{B}]$ while $\varepsilon \to 0$, the exterior dimension $d_x$ in Eq.~(\ref{Eq:codimension}) is determined as:
\begin{equation}
 d_x=D-\alpha,
 \label{Eq:exterior_dimension}
\end{equation}
where $\alpha$ is the uncertainty exponent, also known as {\it exterior codimension}, that can be estimated from log-log plots of $f(\varepsilon)$ as a function of $\varepsilon$ \cite{grebogi1985exterior}.

\begin{figure}[!htp]
 \centering
\includegraphics[width=.48\textwidth]{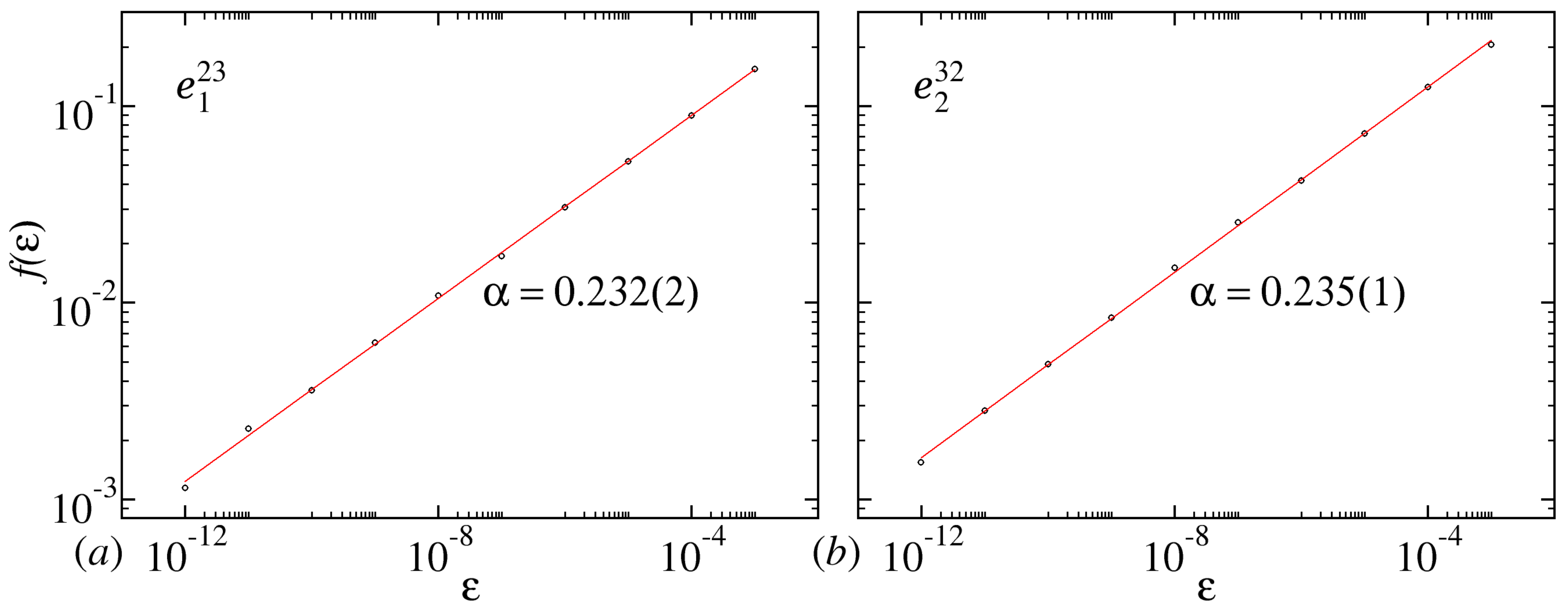}
\caption{Fraction of uncertain parameters $f(\varepsilon)$ as a function of the perturbation $\varepsilon$. A power-law regression $f(\varepsilon)=Ae^{\alpha}$ (red curve) provides the uncertainty exponent $\alpha$ along the extreme curve $e^{23}_{1}$ (a) and $e^{32}_{2}$ (b).}
\label{figure_2}
\end{figure}

{\bf Singular fractal dimension:} Now, we demonstrate the existence of singular fractal sets along periodicity cascades in the parameter space of the map in Eq.(\ref{Eq:log_gauss}). To achieve this, we begin by estimating the uncertainty exponent $\alpha$ of periodicity cascades along the extreme curves $e^{23}_{1}$ and $e^{32}_{2}$. Hence, in Fig.~\ref{figure_2}, we obtain the fraction of uncertain parameters $f(\varepsilon)$ as a function of $\varepsilon$ for parameters along the curves $e^{23}_{1}$ [Fig.~\ref{figure_2}(a)] and $e^{32}_{2}$ [Fig.~\ref{figure_2}(b)]. 
By adjusting a power law function, we find $\alpha = 0.232(2)$ and $\alpha = 0.235(1)$ for the extreme curves $e^{23}_{1}$ and $e^{32}_{2}$, respectively. These values of the uncertainty exponent are significantly different from the ones observed previously in the literature for one-dimensional maps in general ($\alpha \approx 0.41 $) \cite{farmer1985sensitive,grebogi1985exterior,joglekar2014scaling}, or considering only the uncertainty of large attractors ($\beta \approx 0.51 $) \cite{hunt1997structure,joglekar2014scaling}, and continuous-time flows ($\alpha \approx 0.41 $) \cite{medeiros2017sensitive}. Interestingly, the values of the uncertainty exponent $\alpha$ found here contradict the belief that the above values would be universal for one-dimensional maps for a given order of their maxima \cite{farmer1985sensitive, joglekar2014scaling}. Moreover, the lower values of $\alpha$ that we observe along the extreme curves indicate a higher sensitivity to small changes in the parameters on these curves.

\begin{figure}[!htp]
 \centering
\includegraphics[width=.48\textwidth]{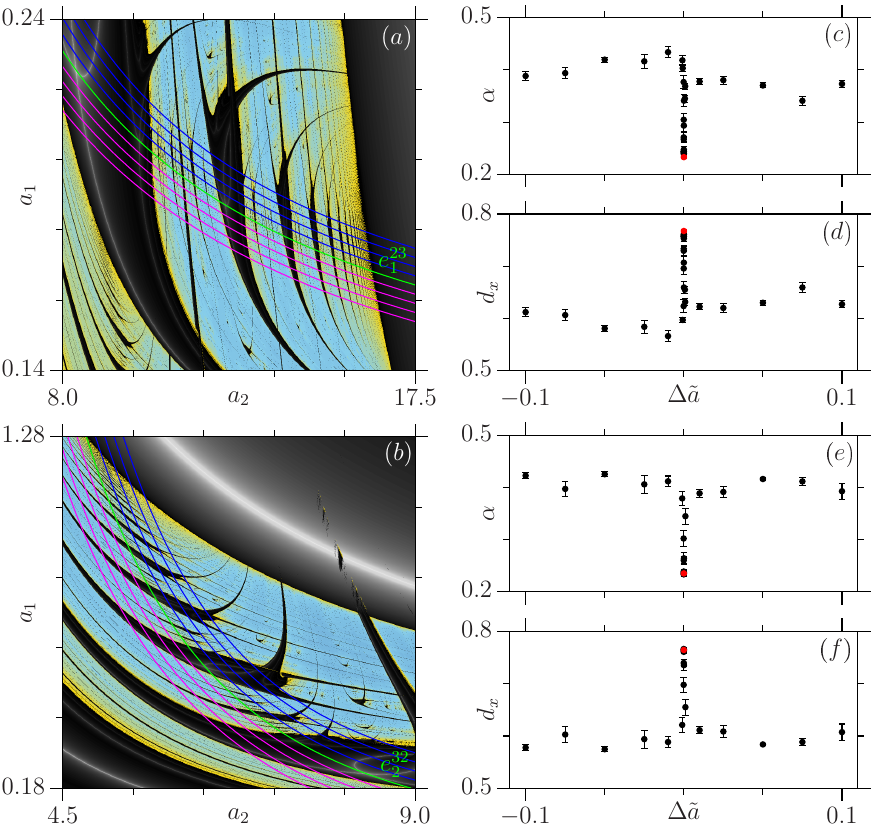}
\caption{(a) and (b) Parameter plane $\Gamma$ illustrating the extreme curve $e^{23}_{1}$ and $e^{32}_{2}$ (green) and parallel curves distant positive (blue) and negative (pink) multiples of $\Delta \Tilde{a} = 0.025$. 
Panels (c) and (e) show the uncertainty exponent $\alpha$ and (d) and (f) present the exterior dimension $d_x$ estimated along the parallel curves within $\Delta \Tilde{a}$ from the curve $e^{23}_{1}$ and $e^{32}_{2}$.
About the picks (red markers) in (c), (d), (e), and (f), $\Delta \Tilde{a} = \pm 10^{-2}, \pm 10^{-3}, \pm 10^{-4}, \pm 10^{-5}, \pm 10^{-6},$ and $\pm 10^{-7}$.}
\label{figure_3}
\end{figure}

Given the significant difference between the uncertainty exponent along the extreme curves and the ones typically observed in the literature, we investigate the values of this exponent in the vicinity of the extreme curves, seeking the transitions towards the typical values of $\alpha$.
Hence, in Fig.~\ref{figure_3}, we calculate $\alpha$ along parallels curves to the extreme curves. 
For this purpose, since the axes of the parameters spaces in Figs.~\ref{figure_3}(a) and (b) have different sizes, we apply the following spatial transformation to normalize these spaces:
\begin{align*}
\quad T:\quad\mathbb{R}^2 &\longrightarrow \quad[0,1]\times [0,1]\\
   (x,y) &\longmapsto \left(\frac{x-x_{m}}{\Delta x}, \frac{y-y_{m}}{\Delta y}\right),
\end{align*}
where $x_{m}$ and $x_{M}$ are, respectively, the $x$ minimum and maximum values in the abscissa axis with size $\Delta x = x_{M}-x_{m}$. 
The same is for the $y$ in the ordinate axis.
Applying $T(a_1, a_2)$ in Eqs. (\ref{Eq:extreme1}) and (\ref{Eq:extreme2}), we represent the extreme curves $e_1^{23}$ and $e_2^{32}$ in the normalized space $[0,1]\times [0,1]$ and we determine parallel curves far $\Delta \Tilde{a}$ from the extreme curve. 
Applying the inverse of the transformation $T^{-1}$ in the parallel curves, we rescue the original coordinate system, as presented in Figs.~\ref{figure_3} (a) and (b).

Following these definitions, we now consider parallel distances $\Delta\Tilde{a} \in [-0.1, 0.1]$ centered on the extreme curves $e^{23}_{1}$  and $e^{32}_{2}$. 
Within this interval, we obtain the uncertainty exponent along curves that are displaced from $e^{23}_{1}$ [Fig.~\ref{figure_3}(c)] and $e^{32}_{2}$ [Fig.~\ref{figure_3}(e)]. 
In addition, we employ Eq.~(\ref{Eq:exterior_dimension}) to calculate the exterior fractal dimension $d_x$ of the sets dwelling along the curves analyzed within $\Delta\Tilde{a}$ [see Figs.~\ref{figure_3} (d) and ~\ref{figure_3} (f)]. 
In the interval surrounding both extreme curves $e^{23}_{1}$ and $e^{32}_{2}$, we observe that the uncertainty exponent $\alpha$, consequently the exterior dimension $d_x$, undergoes abrupt transitions assuming singular values exclusively along the extreme curves located at $\Delta \Tilde{a} = 0$ (red markers).
For the other curves within the interval, $\alpha$ and $d_x$ assume the typical values previously found in the literature.

\begin{figure}[!htp]
    \centering
\includegraphics[width=\linewidth]{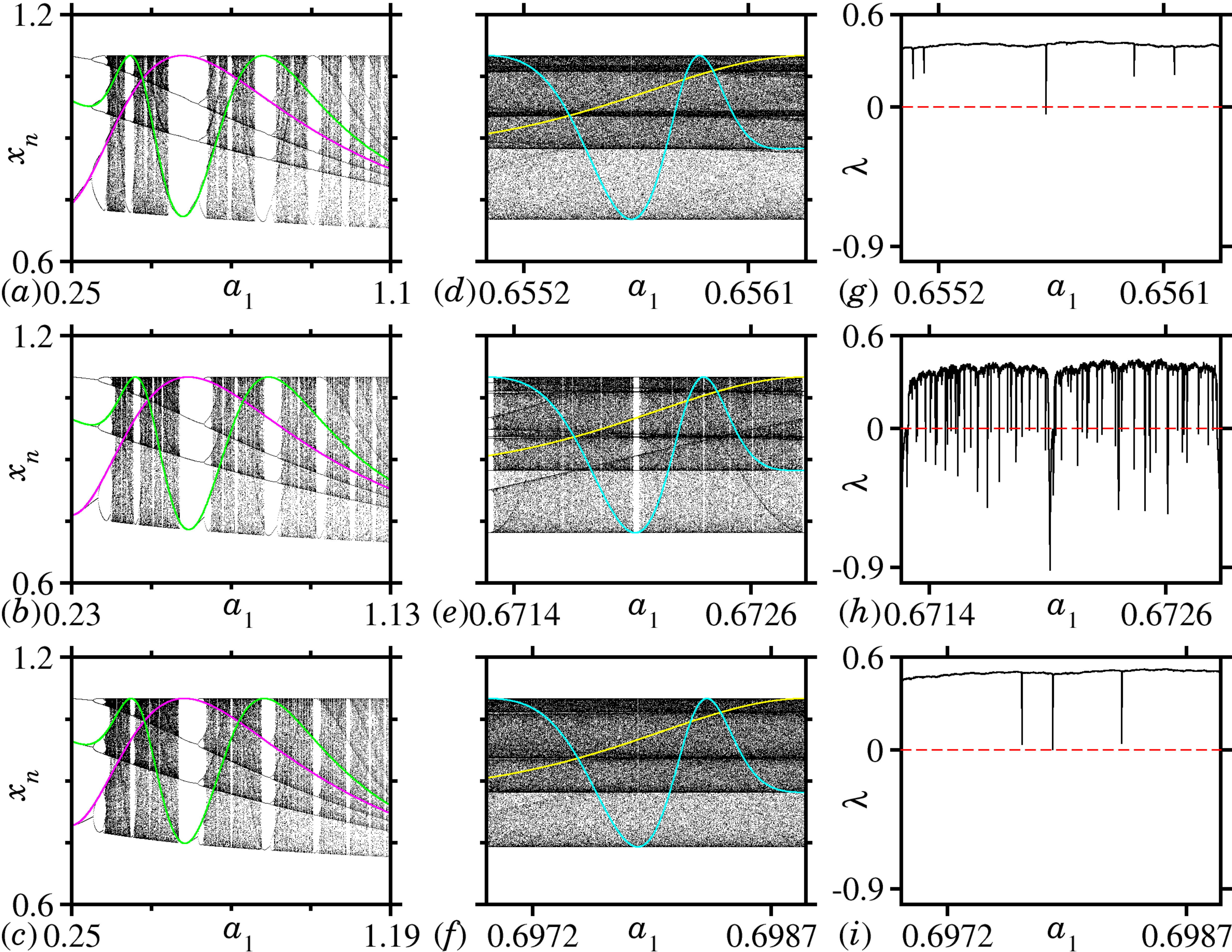}

\caption{(a), (b), and (c) Bifurcation diagrams along curves at $\Delta \Tilde{a} = -0.02$, $\Delta \Tilde{a} = 0$ (extreme curve $e_2^{32}$), and $\Delta \Tilde{a} = 0.02$, respectively. 
(d), (e), and (f) Respective magnification of the bifurcation diagrams shown in the first column. 
(g), (h), and (i) Lyapunov exponents corresponding to the bifurcation diagrams of the middle column.
Colored curves correspond to the $x$ value of the $k$-th iteration of the critical point $x^{\ast}=0$, $f^k(0)$,  at each parameter. $k =$ 5 (pink), 6 (green), 11 (yellow), and 13 (cyan).}
\label{figure_4}
\end{figure}

The results presented in Fig.~\ref{figure_3} demonstrate that periodicity cascades exhibit distinct features along the extreme curves, which give rise to the observed singularities. To elucidate these features, we first recall that the extreme curves intersect the periodic windows at all scales along the periodicity cascade in their superstability locus [as seen in the sequence $(S_i)_{i=1}^\infty$ in Fig.~\ref{figure_1}]. Since these superstability loci serve as the structural foundation for the periodic windows, organizing their extension, it is guaranteed that the extreme curves intersect each window within the periodicity cascades. Conversely, any displaced curve, even though very close to the extreme one, may eventually bypass periodic windows at smaller scales within the parameter space. As a result, the increased density of periodic windows along the extreme curves signifies a more intricate relationship between parameters associated with periodic and chaotic behavior. This leads to a decrease in the uncertainty exponent $\alpha$. Furthermore, the heightened density of periodic windows also populates the one-dimensional curve more densely in a given scale, resulting in the fractal dimension of the boundaries closer to the unity. To better illustrate the increased density of periodic windows, in Fig.~\ref{figure_4}, we present the bifurcation diagrams of the state variable $x$ in Eq.~(\ref{Eq:log_gauss}) along with their corresponding Lyapunov exponent.

In the bifurcation diagrams of Figs.~\ref{figure_4}(a)-\ref{figure_4}(c), the dynamics of $x$ along the extreme curve $e_2^{32}$ [Fig.~\ref{figure_4}(b)] appear similar to those of the displaced curves in Fig.~\ref{figure_4}(a) and Fig.~\ref{figure_4}(c). However, upon closer inspection at a greater scale, as demonstrated in the magnified bifurcation diagrams in Figs.~\ref{figure_4}(d)-\ref{figure_4}(f), the increased density of periodic windows along the extreme curve becomes evident. This higher density of periodic windows can be easily visualized by examining the Lyapunov exponents in Figs.~\ref{figure_4}(g)-\ref{figure_4}(i). Notably, the Lyapunov exponents along the extreme curve [Fig.~\ref{figure_4}(h)] assume negative values very often, in contrast to the displaced curves. The colored curves in all bifurcation diagrams of Fig.~\ref{figure_4} establish a correspondence between the parameter intervals used for the extreme and displaced curves in the magnifications, ensuring comparability among the rows in this figure. These curves correspond to the $k$-th iterate of the critical point $x^{\ast}=0$, denoted as $f^k(0)$ at each parameter. In the pink, green, yellow, and cyan curves, $k$ takes the values of $5$, $6$, $11$, and $13$, respectively.

{\bf Summary:} We employ a class of one-dimensional maps to explore the boundaries between parameters corresponding to self-similar structures of periodic behavior, periodicity cascades, and parameters that lead to chaos, forming fat fractal sets. In two-dimensional parameters spaces of these maps, we estimate the uncertainty exponent and exterior fractal dimension along parameter curves referred to as {\it extreme curves}.
We observe that both measures assume values along these curves that differ significantly from those previously reported in the literature for arbitrary parameter curves, which were believed to be universal. We attribute this difference to a higher density of periodic windows along the extreme curves, which modifies the geometry of the fat fractal sets. Furthermore, in our attempt to identify the transition from the values that we observe along the extreme curves to the typical ones in the literature, we discovered that these transitions occur abruptly near the extreme curves. Based on this observation, we propose that the uncertainty exponent and, consequently, the exterior fractal dimension exhibit singular behavior along the extreme curves.

\begin{acknowledgments}
E.S.M acknowledges the support by the Deutsche Forschungsgemeinschaft (DFG) via the project number 454054251 (FE 359/22-1). R.O.M.T. is in debit with Sebastian van Strien, Dmitry Turaev, and Jeroen Lamb for the reception and illuminating discussions at Imperial College London in the end of 2024 and acknowledges the partial support of National Council for Scientific and Technological Development – CNPq, project number 408522/2023-2. C.E.P.A. and J.D.V.H. thanks Federal Institute of Education, Science and Technology of
South of Minas Gerais - IFSULDEMINAS.
\end{acknowledgments}


\end{document}